\documentclass[%
 twocolumn,
nofootinbib,
 amsmath,amssymb,
 aps, physrev,
]{revtex4-2}

\usepackage{graphicx}
\usepackage{dcolumn}
\usepackage{bm}
\usepackage{hyperref}
\usepackage{xcolor}

\usepackage{subcaption}

\usepackage{soul}
\usepackage{float}

\begin{document}

\preprint{APS/123-QED}

\title{\textbf{History of UHECR production in Centaurus A}
}%

\author{Cainã de Oliveira}
  \email{Contact author: olivcaina@gmail.com}
\author{Vitor de Souza}%

\affiliation{%
 Sao Carlos Institute of Physics, University of Sao Paulo, IFSC – USP,
13566-590, Sao Carlos, SP, Brazil.
}%

\date{\today}

\begin{abstract}
The origin of the UHECR continues to puzzle, however, an excess of detection in the direction of the radio galaxy Centaurus A (Cen~A) raises the possibility of this object being the first UHECR source identifiable. Cen~A is known to be currently active, and also exhibits known past episodes of high activity. In this work, we investigate whether the known activity episodes in Cen A may be related to the excess events in the \textit{Centaurus region}. Analysing the energy of the events and the overall mass composition of UHECR, we report that an activity in the last $\sim30$~Myr is necessary to explain the excess of events. This period perfectly fits with the timescale where the transition regions and the Giant Lobes must be energized, as revealed by radio and $\gamma$ ray observations.

\end{abstract}

\maketitle


\section{\label{sec:intro} Introduction}

The origin of ultra-high-energy cosmic rays (UHECR) remains unknown for more than half a century~\cite{linsley1963}. Charged particles are deflected by the Galactic and extragalactic magnetic fields, making it difficult to trace them back to their sources. Deflections are smaller for higher-energy particles from nearby sources. Moreover, the most energetic particles likely originate within $\sim$100 Mpc, as energy losses from interactions with the cosmic microwave and extragalactic background light limit their travel distance~\cite{gzk:g,gzk:zk,lang2020revisiting}. This supports the prospect of conducting UHECR astronomy focused on nearby sources~\cite{cronin:uhecrastronomy}.

The possibility of UHECR acceleration in radio galaxies, and particularly Cen~A, has been discussed before~\cite{ROMERO1996279,james_fornax,rieger2022active}, and it gained strength with the detection of the excess of events~\cite{liu2012excess,deOliveira_2022,deOliveira_2023}. Cen~A proximity ($3.8 \pm 0.1$~Mpc), angular size ($\sim 9^\circ$), and relative orientation allow the detection of high-quality multiwavelength data, revealing its history and spatial structures~\cite{mckinley2022multi,hess2020resolving,sun_2016,croston_2009,israel1998centaurus}.
    
Despite experimental and theoretical indications, recent analysis~\cite{james_backflow,james_fornax} suggest that the current jet activity of Cen~A seems to satisfy the power constraints necessary to accelerate UHECRs only for highly ideal scenarios. However, a powerful past activity can not be ruled out, and in this case, the radio lobes will now act as UHECR reservoirs~\cite{james_backflow,james_fornax}. In this work, we explore the history of Cen~A to constrain which past activity may be responsible for the UHECR detected today. We identify the time window during which UHECRs were accelerated in the inner structure of Cen~A and remained confined in its lobes for long enough to be detected on Earth today. We demonstrate that the energization timescale of the Giant Lobes of Cen~A is consistent with the production of UHECRs that contribute to the hotspot observed by the Pierre Auger Observatory, taking into account both composition and energetic requirements.

\section{Time constraints of acceleration of particles in Cen A}~\label{sec:time}

In this section, we calculate the time between the scape of the UHECR from the acceleration region in Cen~A and the detection on Earth. The calculated time is compared to the history of events and age of structures in Cen A, as well as to the data measured by the Pierre Auger Observatory.

The radio structure of Cen~A reveals the presence of three extended regions around the central galaxy, known as Giant/Outer lobes~\cite{israel1998centaurus}, Northern Middle Lobe, and Inner Lobes. The Giant Lobes (Northern and Southern) have size larger than $480$~kpc and dynamic age of $\sim 1.6$~Gyr~\cite{mckinley2022multi}. The radio and $\gamma$-ray emissions limits the energy supply to no longer than $\sim 30$~Myr ago~\cite{hardcastle_lobes,Eilek_2014}. Each Inner Lobes is $\sim 5$~kpc long  and $\sim (1 - 2)$~Myr old~\cite{Neff_2015paper1}.  The Northern Middle Lobe (referred to as Middle Lobe in this study) is the northern transition region connecting the Inner to the Giant Lobes. It is characterized by aligned radio and X-ray knots and filaments that resemble a large-scale jet. Its origin has been proposed as the interaction of a broad, large-scale outflow (from the AGN or starburst activity) boosted by an AGN jet outburst, with cold and warm gas clouds, producing a region of turbulence and inducing star formation~\cite{Neff_2015paper2,mckinley2022multi}. The outflow properties derived by \citet{mckinley2022multi} suggest an age of at least $56$~Myr to account for the X-ray power emitted from knots in the Middle Lobe. The energy flow through the transition regions is necessary to energize the Giant Lobes, corroborating a re-energization event no more than $\sim30$~Myr ago. The presence of short-lived structures requires re-energization on $\lesssim 10$~Myr~\cite{Neff_2015paper1,Neff_2015paper2}. Ongoing starburst activity is also reported to Cen~A. Its age is $\gtrsim 50$~Myr, with a recent star formation in the last $2-50$~Myr~\cite{Neff_2015paper2}. These timescales and sizes will be used to compare with the calculations below.

To connect the history of activity in Cen~A with the UHECR signal currently measured at Earth, it is necessary to account for the time delays to which UHECRs are subject. We assume that any of the internal structures in Cen~A (jet, Inner Lobes, or transition regions) accelerate UHECR. After being injected from the acceleration site, the time necessary to detect the particles is the sum of the time to escape the source ($\tau_{\rm esc}$), and time delays from the extragalactic and Galactic propagation ($\delta t_{\rm prop}$).

The time necessary for particles to escape the source can be estimated by a combination of microscopic propagation regimes and advection in the Giant Lobes. Given the age and size of the Giant Lobes, UHECRs possibly accelerated in any of the internal structures will need to cross them before escaping to the interstellar medium. In the case where acceleration occurs in the Giant Lobes, the constraints derived below can be seen as upper limits, since particles do not necessarily need to cross the entire Lobe extension to escape.

Lobes of radio galaxies are filamentary and magnetized environments~\cite{sullivan_lobes,wykes_2014_lobes}. The time necessary for escape from the Giant Lobes can be estimated by a combination of microscopic propagation regimes and advection of particles in the ambient plasma. During the propagation throughout the Giant Lobes, energy losses could modulate the energy spectrum. However, as shown in Appendix~\ref{app:energy_loss}, energy losses inside the Giant Lobes can be neglected for the energy range we are considering.

The microscopic propagation involves the transition between rectilinear and diffusive regimes. We model the lobes of Cen~A as a turbulent, approximately spherical medium. As shown by~\citet{Aloisio_2009}, in the case of isotropic turbulence and injection, these regimes can be fully described by the generalized propagator
\begin{equation} 
    \label{eq:prop_juttner}
    P_J(r,t,E) = \frac{\Theta \left( ct - r \right)}{4\pi (ct)^3} \frac{\alpha(t,E)}{K_1(\alpha(t,E))} \frac{e^{-\alpha / \sqrt{1 - (r/ ct)^2}}}{[1 - (r/ ct)^2]^2},
\end{equation}
where $K_1$ is the Bessel function and $\alpha(t, E) = c^2 t/ 2 D(E)$, in the absence of energy losses.

For a rectilinear propagation, particles take $\tau_{\rm rect} = L / c$ to escape a region of size $L$. The same distance is travelled in a timescale $\tau_{\rm dif} = L^2 / 6 D(E)$ in the diffusive regime. Considering eq.~\ref{eq:prop_juttner}, a general timescale to cross $L$ can be obtained by writing
\begin{equation}
    \tau_{\rm esc}(E) = \frac{L^2}{\langle r^2 \rangle(L/c,E)} \frac{L}{c},
\end{equation}
where $\langle r^2 \rangle (t,E) = \int r^2 P_{J}(r,t,E) d^3 r$, and we use the definition $D = \langle r^2 \rangle / 6 t$. Along this work, we adopt the diffusion coefficient found by~\citet{harari_dif_coef}
\begin{equation}
    D(E) = \frac{c}{3}\ell_c \left[ 4 \left(\frac{E}{E_c}\right)^2 + 0.9 \frac{E}{E_c} + 0.23 \left(\frac{E}{E_c}\right)^{1/3} \right]
\end{equation}
where $E_c = 0.9 \left( B_{\rm rms} / \mu{\rm G} \right) \left( \ell_c / {\rm kpc} \right)$~EeV, $B_{\rm rms}$ is the root mean square of the magnetic field intensity, and $\ell_c$ its coherence length. We assume a Kolmogorov spectrum for the magnetic field turbulence. Radio and $\gamma$-ray measurements from the Giant Lobes indicate $B_{\rm rms} \approx 1~\mu$G~\cite{PAPER_cenA,sun_2016,hardcastle_lobes,Eilek_2014}. Observations of radio filaments in the southern Giant Lobe suggest the scale $\ell_{\rm max} \sim 50$~kpc~\cite{wykes_2015_lobes,Eilek_2014}, implying $\ell_c \sim 10$~kpc for a Kolmogorov turbulence~\cite{Harari_2002}.

Advective motions can play an important role in the escape from lobes. As discussed by \citet{james_taylor_flickering}, a reliable estimation for the advection timescale $\tau_{\rm adv} = L / v_{\rm adv}$ is hard to obtain. For Cen~A, a large-scale outflow with speed $\sim1100~{\rm km~s}^{-1}$ is measured~\cite{mckinley2022multi}. Despite the outflow should thermalize at a radius $\sim75$~kpc~\cite{mckinley2022multi}, we take the pragmatic approach of using its speed to estimate the advection timescale.

The exact value of $L$ depends on the exact structure and time at which the acceleration happened. Based on the size of the Giant Lobes and the spherical approximation, we consider $L \sim 200$~kpc.

\begin{figure}[t]
\includegraphics[width=0.8\linewidth]{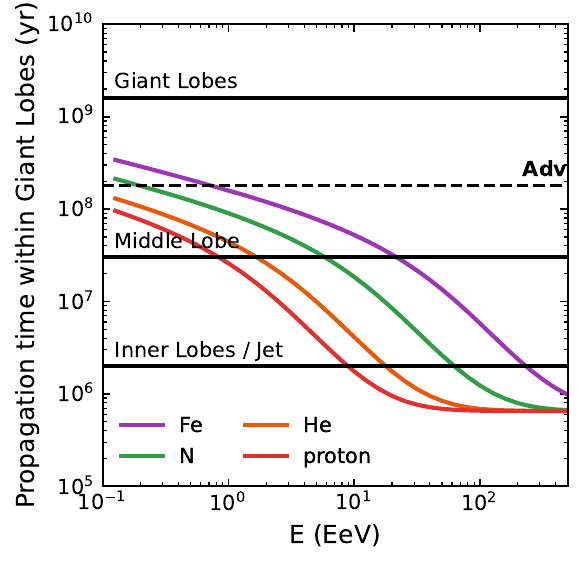}
\caption{\label{fig:timescales_new} Time necessary for UHECR to escape the Giant Lobes of Cen~A for different UHECR species. The age and timescales for different structures in Cen~A (continuous black lines) and the advection timescale inside the lobes (black dashed line indicated by Adv) are also shown. The Middle Lobe timescale is taken as $30$~Myr.}
\end{figure}

Figure~\ref{fig:timescales_new} shows the time necessary for UHECR to escape the Giant Lobes of Cen~A for different UHECR species. UHECRs of higher energy realize rectilinear trajectories, and escape the lobes after $\sim0.6$~Myr. However, diffusion starts to be important for lower energies ($\lesssim10$~EeV for protons and $\lesssim100$~EeV for Nitrogen), causing a significant delay in the escape time. In general, advection is only important below $\sim1$~EeV and can be neglected for our purposes. Considering the age of the current activity, only UHECR with energies above $\sim10-100$~EeV have time enough to escape the Giant Lobes. For particles possibly accelerated during the Middle/Giant Lobes re-energization timescale, only protons with energies $\gtrsim1$~EeV and Fe with energies $\gtrsim 20$~EeV escaped.

After escaping the Giant Lobes, particles will propagate through the extragalactic environment before hitting Earth. Extragalactic magnetic fields delay the arrival time of UHECR compared to a rectilinear propagation. \citet{silvia_cenA} estimates an increase $\delta t_{\rm prop} \sim \theta_{\rm rms}^2 d / 6c$. To estimate $\delta t_{\rm prop}$ for UHECR within the \textit{Centaurus} hotspot, we take $d = 3.8$~Mpc~\cite{harris2010distance} and the scattering angle as the hotspot radius $\theta \sim 27^\circ$~\cite{supergalactic_auger}, the time delay of UHECR populating the hotspot is $\delta t_{\rm prop}\sim0.5$~Myr. Despite that, larger time delays were proposed by \citet{mbarek2025propagationdelaysultrahighenergycosmic}, depending on particles' rigidity and the coherence length of the extragalactic magnetic field.

The timescales obtained proved that UHECR injected by Cen~A are likely to remain imprisoned inside the Giant Lobes. The residence time is dependent on the energy and charge of the particles, varying from $\sim 0.6$~Myr for particles performing a rectilinear propagation to $\sim 100$~Myr for less energetic diffusing particles. These timescales are likely to dominate over the time delay caused by the extragalactic propagation ($\sim 0.5$~Myr) in the limit of small scattering. Only light (proton, helium) particles with energies above $\sim 10$~EeV accelerated from the current activity of Cen~A had time to escape from the Giant Lobes when the current jet activity is considered. UHECR accelerated during the period of re-energization timescale of the Middle Lobe are more likely to contribute to an intermediate/heavy composition measured at Earth for energies above $\sim10$~EeV.

In the next section, the imprisoning effect in the Giant Lobes will be used to constrain the episodes of activity that could be contributing to UHECR currently measured at Earth.


\begin{figure}[b]
  \centering
  \begin{subfigure}{0.49\textwidth}
    \includegraphics[width=0.8\linewidth]{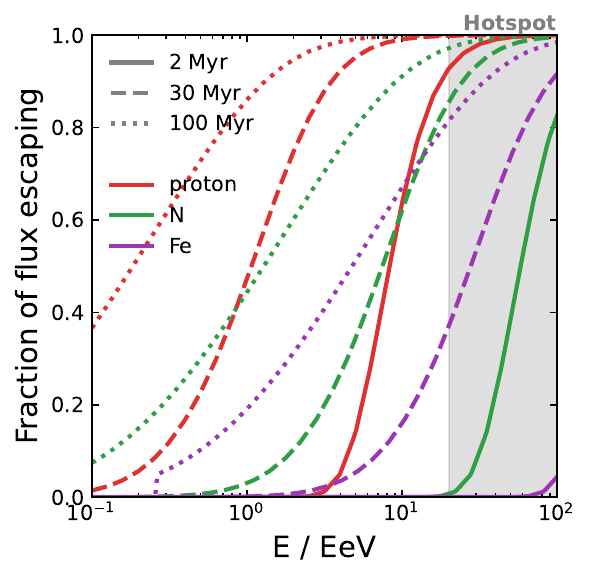}
  \end{subfigure}
  \hfill
  \begin{subfigure}{0.49\textwidth}
    \includegraphics[width=0.8\linewidth]{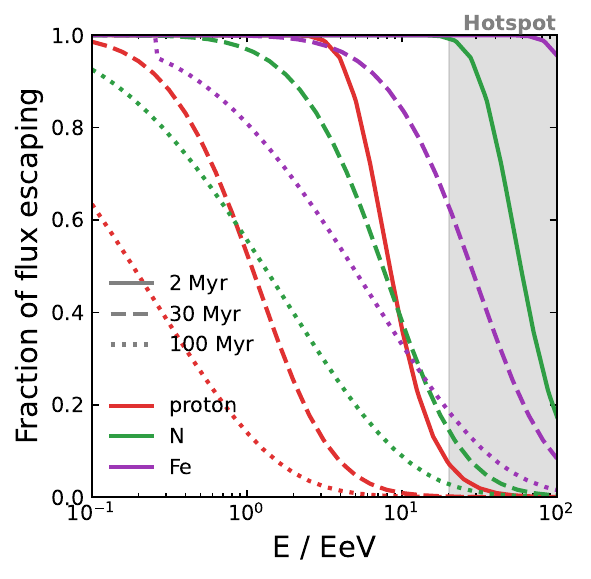}
  \end{subfigure}
  \caption{UHECR flux fraction escaping from the Giant Lobes for different injection scenarios. Proton, N, and Fe species are considered. Top panel: Injection starting at $2$~Myr (continuous lines), $30$~Myrs (dashed lines), and $100$~Myrs (dotted). Bottom panel: Injection ceasing $2$~Myr (continuous), $30$~Myrs (dashed), and $100$~Myrs (dotted) ago. The energy range where the excess in the \textit{Centaurus region} is reported~\cite{supergalactic_auger} is shown as a gray band.}
  \label{fig:fraction_cenA}
\end{figure}

\section{When the UHECR of the Centaurus hotspot was generated ?}~\label{sec:when}

The imprisoning of particles in the Lobes will result in composition-dependent suppressions of the flux according to the particle acceleration age. Considering a point source, the density at a position $r$ is given by the convolution of $P_J$ with the time injection $Q(t)$,
\begin{equation}
    n(r,t,E) = \int_{-\infty}^\infty d\tau~ Q(\tau) P_J(r,t - \tau,E).
\end{equation}

To constrain the episode of high activity where particles should be accelerated in Cen~A, consider that the UHECR injection starts or ceases at a time $t_0$. It can be modeled as Heaviside step functions, $\Theta(t - t_0)$. In these cases, the particle density is given by
\begin{equation} \label{eq:ID1}
    n(r,t,E) = \int_{t_0}^\infty d\tau P_J(r,t - \tau,E) = \int_{-\infty}^{t-t_0} d\tau P_J(r,\tau,E).
\end{equation}

The flux leaving the source can be obtained via continuity equation, $\nabla \cdot \mathbf{j} = - \partial_t n$, which can be integrated for spherical symmetry,
\begin{equation} \label{eq:flux}
    J(R,t,E) = 4 \pi R^2 j = 4 \pi \int_{R}^{c t} dr~r^2 \partial_t n(r,t,E),
\end{equation}
that represents the total flux of particles with energy $E$ leaving the source region of radius $R$ at time $t$. Using Eq.~\ref{eq:ID1}, the flux leaving the source is given by\footnote{Due to numerical instabilities, the solution is combined with a truncated Gaussian for lower energies/long times, as in \citet{lang2020revisiting}.}
\begin{equation} ~\label{eq:escaping_flux}
    J(R,t,E) = 4 \pi \int_{R}^{c t} dr~r^2 P_J(r,t-t_0,E).
\end{equation}

Figure~\ref{fig:fraction_cenA} presents the flux escaping the Giant Lobes, considering that acceleration has been occurring for $\Delta t =2,~30,~100$~Myr (top panel) or ceased for these periods (lower panel). A strong cutoff, resulting from the diffusion time, is observed in the flux according to the source activity age and UHECR rigidity. For UHECR injection starting recently, the lightest particles are the first to escape, imposing a lower limit to source activity age. Only protons can escape the Giant Lobes for the current AGN activity ($\sim2$~Myr). Nitrogen nuclei with energies $\sim20$~EeV take at least $30$~Myr to escape, while the situation is more dramatic for Fe nuclei, where a significant fraction of events above $\sim20$~EeV can escape only for an acceleration larger than $\sim100$~Myr. The conclusion is reversed if the UHECR injection stopped years ago. Heavier particles are the last to escape, imposing an upper limit on the source activity age.

The excess in the \textit{Centaurus region} extends from high energies ($>60$~EeV) down to $\sim20$~EeV. In this energy range, there are indications for an intermediate composition, probably Nitrogen-like~\cite{AbdulHalim:2023Yd}. Composition-dependent anisotropies points to an intermediate composition as well~\cite{mass_dependent_anisotropy}. Based on Figure~\ref{fig:fraction_cenA}, Nitrogen nuclei with energies above $\sim10$~EeV require the source to be active for the last $\sim 30$~Myr. For Iron nuclei of $\sim20$~EeV to escape the source, the injection must not cease earlier than $\sim30$~Myr. Note that if UHECRs are not injected in the Lobes for more than $\sim 100$~Myr, the UHECR flux above $10$~EeV will be highly suppressed. These timescales indicate that an activity for a period $\gtrsim2$~Myr and not ceased later than $\sim30$~Myr is necessary to accelerate UHECR, if the central engine of Cen~A is responsible for the observed UHECR hotspot.

The results above suggest that a past activity $\sim2-30$~Myr ago may have accelerated UHECR that are now being detected. This timescale agrees with the re-energization timescale of the Middle and Giant Lobes. The scenario of a recent $\lesssim30$~Myr activity combined with a short period of low activity is expected for Cen~A. However, a long $\sim50-100$~Myr quiet period is highly unlike~\cite{Neff_2015paper2}.

\subsection{Constraints from energy spectrum and composition}

Based on the previous discussion, the propagation through the lobes should considerably modify the energy spectrum and composition of UHECR from Cen~A. To illustrate this effect, consider a commonly adopted power-law energy spectrum with an exponential cutoff at energy $Z_j R_{\rm max}$ and spectral index $s$,
\begin{equation}
    \frac{dN}{dE} \propto \sum_j f_j E^{-s} e^{E / Z_j R_{\rm max} }.
\end{equation}
From Equation 19 of \citet{james_backflow} we estimate $R_{\rm max} \sim 5-10$~EeV applying the power estimates for Cen~A jet $\sim2-6 \times10^{43}~{\rm erg~s}^{-1}$ from \citet{Neff_2015paper1}. For illustrative purposes, we take $s=2$ and assume a rectilinear extragalactic propagation from Cen~A. As shown in Appendix~\ref{app:energy_loss} (Fig.~\ref{fig:timescales_cenA}), in this case, extragalactic propagation causes significant changes in the spectrum only for Helium and Nitrogen nuclei with energies above $\sim100$~EeV, where the energy spectrum is suppressed by the exponential cutoff ($E_{\max} = 20$ and $70$~EeV, for He and N, respectively). Therefore, the extragalactic propagation will be ignored in the following analysis. It keeps the effect of the propagation inside the lobes clearer.

\begin{figure*}[b]
\includegraphics[width=0.75\linewidth]{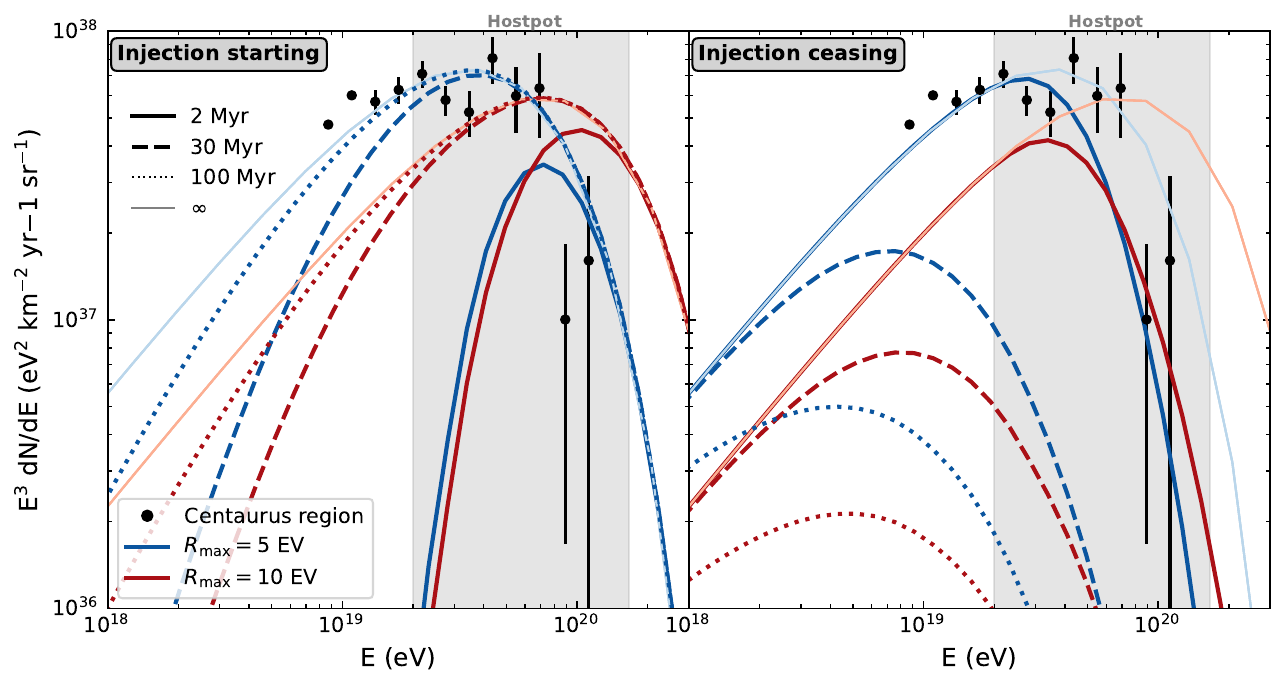}
\caption{\label{fig:spectrum_N_cenA} Energy spectrum of Nitrogen nuclei escaping the lobes of Cen~A for several temporal injections.
Injection starting (left) or ceasing (right) in $2$, $30$, and $100$~Myr are shown. Two maximum rigidities, $5$ (blues) and $10$~EeV (reds), are considered. For comparison, the energy spectrum measured in a window $20^\circ$ around Cen~A is also shown~\cite{AbdulHalim_2024}. The energy spectrum of continuous infinity injection ($\infty$) is arbitrary normalized, and corrected by the fraction of escaping flux (eq.~\ref{eq:escaping_flux}).}
\end{figure*}

For comparison with the energy range of events of the \textit{Centaurus region}, we consider an intermediate composition. As stated in the previous section, a pure Nitrogen composition will be considered. Figure~\ref{fig:spectrum_N_cenA} shows the energy spectrum escaping the lobes for the emission starting or stopping at $2,~30,~100$~Myr. For a qualitative discussion, the energy spectrum reported by the Pierre Auger Collaboration within the \textit{Centaurus region} is also presented. In this case, the high-energy events in the \textit{Centaurus region} cannot be explained if the acceleration has been stopped for $\sim30$~Myr for a Nitrogen-like composition.

\begin{figure*}
  \centering
  \includegraphics[width=0.75\linewidth]{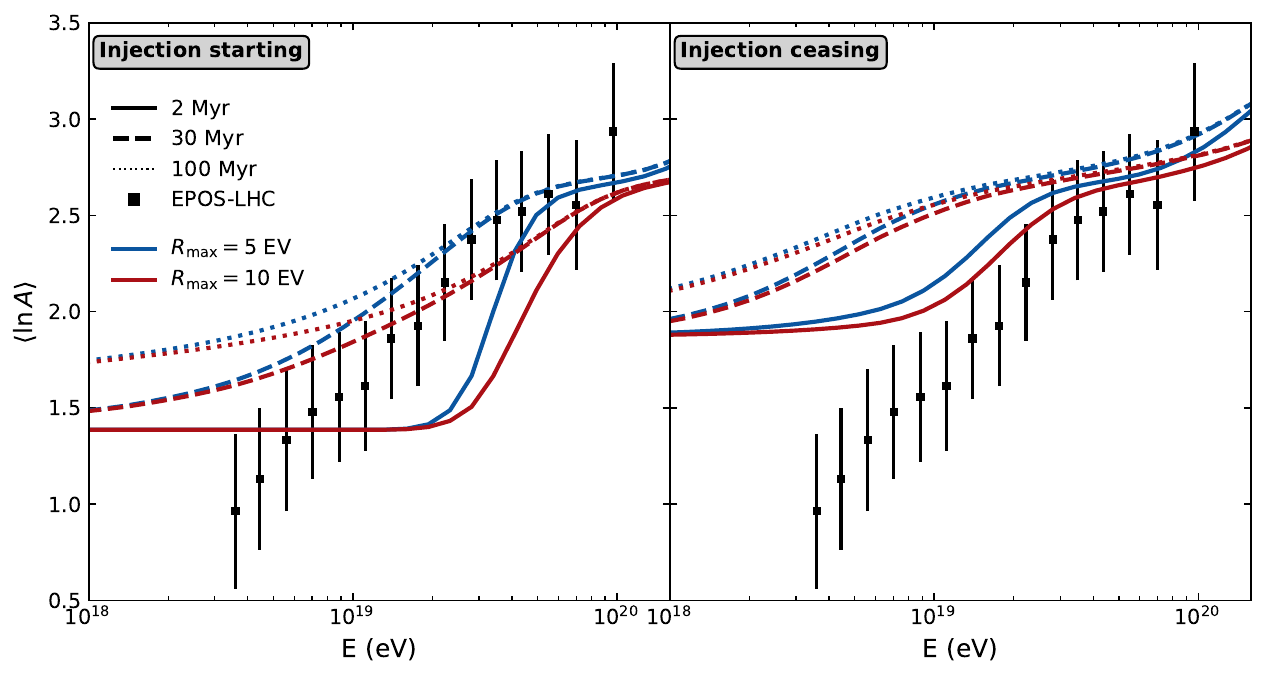}
  \caption{Evolution of the mean composition with the energy escaping the Giant Lobes for several time injections, considering a Wolf-Rayet-like composition. The mean logarithm of the composition obtained by the Pierre Auger Collaboration~\cite{mass_deep_learning} for the EPOS-LHC hadronic model is shown for comparison.}
  \label{fig:composition_CenA}
\end{figure*}

The results found in Figure~\ref{fig:fraction_cenA} point to a composition highly dependent on the age of the acceleration process. To illustrate the mass evolution with the energy according to the injection time, we assume a composition inspired by Wolf-Rayet stars~\cite{liu2012excess} ($f_{\rm H}=0$, $f_{\rm He}=0.62$, $f_{\rm N}=0.37$, $f_{\rm Si}=0.01$, $f_{\rm Fe}=0$), as suggested by \citet{muller}. Figure~\ref{fig:composition_CenA} shows the evolution of $\langle \ln A \rangle$ with energy for different starting and stopping injection times for Cen~A and the two $E_{\rm max}$ estimated above. The mean mass evolution reconstructed by the Pierre Auger Observatory~\cite{mass_deep_learning}, considering the EPOS-LHC hadronic model, is also shown. In the absence of energy losses, $\langle \ln A \rangle$ is independent of the spectral index $s$ for the energy spectrum we assume. If the source stopped the injection for a period $\gtrsim2$~Myr, the composition measured on Earth will be mainly heavy, in disagreement with the data. For the composition trend to be compatible with the data is necessary that an injection starting less than $100$~Myr, with better agreement for timescales $\lesssim30$~Myr. This result is consistent with our previous discussion about the timescales where the acceleration of particles in Cen~A must occur.

\section{Conclusion}

In this paper, we place stringent constraints on the time window during which the acceleration of ultra-high-energy cosmic rays (UHECRs), currently observed at Earth and potentially originating from Cen~A, must have occurred. Comparing the energy dependency of the \textit{Centaurus excess}, we obtained that UHECR acceleration in Cen~A must have happened $\sim 2-30$~Myr ago. A particle acceleration occurring during the recent ($\sim2$~Myr) reactivation of the AGN that produced the current jet and Inner Lobes cannot explain the energy spectrum measured in the \textit{Centaurus excess} region when an intermediate, Nitrogen-like composition is considered. An injection of UHECR in a distant past activity, $\gtrsim100$~Myr, is constrained by energetic and composition arguments. This indicates that the formation of the Giant Lobes ($\sim$~Gyr) cannot account for the high-energy events detected.

If Cen~A were the dominant source of ultra-high-energy cosmic rays (UHECRs) at extreme energies, injecting primarily light or intermediate-mass nuclei\cite{silvia_cenA,biermman&desouza}, one would expect a detectable clustering of events with energies $\gtrsim 100$~EeV in its direction. However, the escape time for UHECRs from the Giant Lobes at these energies is approximately 0.6~Myr (see Figure~\ref{fig:timescales_cenA}). A temporary reduction in the source’s activity over this timescale would be sufficient to suppress the current detection of such extreme events from Cen~A.

In this scenario, only heavier nuclei -- potentially confined within the lobes for extended periods (on the order of $\sim2$~Myr) -- could escape. These nuclei experience greater deflections due to their higher charge in both Galactic and extragalactic magnetic fields, which can lead to the observed isotropy at the highest energies. Consequently, the absence of a significant excess of events with energies $\gtrsim 100$~EeV in the direction of Cen~A can be explained by a relatively short ($\sim0.6-2$~Myr) quiescent phase — consistent with the episodic nature of Cen~A's activity\cite{Neff_2015paper2} — or by limitations in its ability to accelerate particles beyond this energy. In both cases, the particles escaping the lobes at the highest energies are expected to be predominantly heavy nuclei.

This suggests that the current activity of Cen~A is unlike to produce the events detected in the hotspot as well. As we have shown in a previous work~\cite{deOliveira_2022}, heavy Fe-like species can suffer strong deflections due to the EGMF and GMF.

The observed energy and composition of the hotspot suggest acceleration timescales on the order of $\sim 2–30$~Myr, aligning well with the recent re-energization episodes of the Middle and Giant Lobes of CenA. This temporal consistency strongly supports the hypothesis that these structures are responsible for the observed excess. Given the established evidence of past activity in Cen~A, attributing the \textit{Centaurus} excess to this source is both plausible and well supported by current observational data

\begin{acknowledgments}
CdO thanks Henrique Malavazzi and Leonardo Paulo Maia for useful discussions. This study was financed, in part, by the São Paulo Research Foundation (FAPESP), Brasil. Process Number 2025/03325-5, 2021/01089-1, 2020/15453-4 and 2019/10151-2. The authors acknowledge the National Laboratory for Scientific Computing (LNCC/MCTI, Brazil) for providing HPC resources of the SDumont supercomputer, which have contributed to the research results reported within this paper. URL: \url{http://sdumont.lncc.br}
\end{acknowledgments}

\bibliographystyle{apsrev4-2}
\bibliography{apssamp}

\appendix

\section{Energy losses inside the Lobes} \label{app:energy_loss}

During propagation, the energy spectrum can be modified by hadronic and photohadronic interactions. Modulations of the spectrum will be relevant if the interaction and escape timescales are comparable, $\tau_{\rm int} \gtrsim \tau_{\rm esc}$. We consider photohadronic interactions with the cosmic microwave background and extragalactic background light~\cite{gilmore_2012_EBL} for redshift $z=0$, obtained using the {\fontfamily{cmtt}\selectfont CRPropa} code~\cite{batista2016crpropa}. 

Hadronic (proton-proton or nucleus-proton) interactions can occur with the thermal material of the lobes of density $n_{th}$. Since the proton-proton cross section is weakly dependent on energy, we follow previous works writting $\tau_{\rm pp} \approx 1.7 \times 10^{6} n^{-1}_{-4}$~Myr, where $n_{th}= 10^{-4} n_{-4}~{\rm cm}^{-3}$~\cite{aharonian_pp,fraija_lobes,hardcastle_lobes}. Taking $n_{-4} \sim 0.1-1$~\cite{mckinley2022multi} results in $\tau_{\rm pp} \sim 10^6 - 10^{7}$~Myr. For a nucleus of mass $A$, the cross section and the timescale increases by $A^{3/4}$, a factor $\sim20$ for $^{56}{\rm Fe}$.

\begin{figure*}[t]
\includegraphics[width=0.8\linewidth]{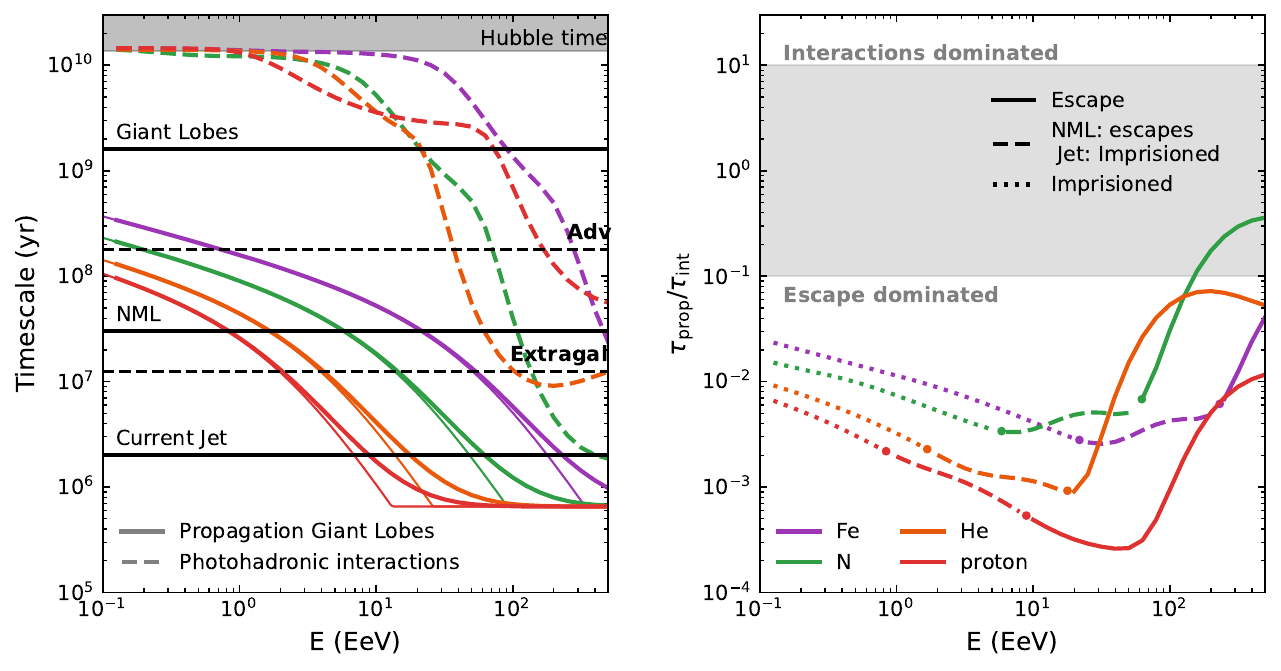}
\caption{\label{fig:timescales_cenA} Timescales for Cen~A. Left panel: Timescales for propagation of different UHECR species (continuous colored line) in the Giant Lobes are shown altogether with timescales for different structures in Cen~A (continuous black lines). Interaction timescales for the CMB and EBL are shown in dashed lines. Black dashed lines are estimations for the extragalactic time interval necessary UHECR to reach Earth, considering a rectilinear trajectory (Extragal), and the advection timescale inside the lobes (Adv). Together with the propagation time, thin continuous lines represent the propagation timescale for rectilinear and diffusive regimes. Right panel: Ratio between propagation and interaction timescales for different UHECR species. Continuous lines indicate the fraction of UHECR expected to escape from the Giant Lobes, both if generated in a Middle Lobe re-energization (NML, $\sim30$~Myr) or the current jet timescale ($\sim2$~Myr). Dashed lines indicate escape if accelerated on Middle Lobe re-energization, but not if injected during the current activity. Dotted lines indicate the fraction that remains imprisoned. The gray band represents the transition between a regime escape- or interaction-dominated.}
\end{figure*}

Figure~\ref{fig:timescales_cenA} presents $\tau_{\rm esc}$, $\tau_{\rm int}$, and $\tau_{\rm adv}$ for different species (proton, $^4$He, $^{14}$N, $^{56}$Fe), compared with the ages of different structures in Cen~A (left panel). To determine the importance of interactions over the energy spectrum, the ratio $\tau_{\rm esc} / \tau_{\rm int}$ is also shown (right panel). In general, $\tau_{\rm int} / \tau_{\rm esc} \ll 1$, indicating that modulations of the energy spectrum by interactions inside the Giant Lobes are unlike. 

Hadronic interactions take place on timescales much longer than needed for escape, and are not shown. UHECR of high energy realize rectilinear trajectories, and escape the lobes after $\sim0.6$~Myr. However, diffusion starts to be important for lower energies ($\lesssim10$~EeV for protons and $\lesssim100$~EeV for Nitrogen), causing a significant delay in the escape time. In general, advection is only important below $\sim1$~EeV and can be neglected for our purposes.

\end{document}